\documentclass[aps,
prd,
reprint,
floatfix,
superscriptaddress,
longbibliography,
nobalancelastpage
]{revtex4-2}

\usepackage{graphicx}
\usepackage{physics}
\usepackage{booktabs}
\usepackage{amsthm,amssymb,mathrsfs}
\usepackage{tensor}
\usepackage{bm}
\usepackage[caption=false]{subfig}

\usepackage{feynmp-auto}

\usepackage{hyperref} 
\hypersetup{colorlinks, allcolors=blue}

\usepackage{xargs} 
\usepackage{xcolor}
\usepackage{slashed}

\usepackage[inline]{enumitem}

\usepackage{enumitem}
\setlist[itemize]{leftmargin=*}
\setlist[enumerate]{leftmargin=*}

\let\temp\phi
\let\phi\varphi
\let\varphi\temp

\providecommand{\calL}{\mathcal{L}}

\newcommand{\iu}{\mathrm{i}} 

\newcommand{\D}{\mathrm{d}} 
\newcommand{\sign}{\operatorname{sgn}} 

\let\v\relax %
\newcommand{\v}[1]{\ensuremath{\mathbf{#1}}} 

\makeatletter
\newsavebox{\@brx}
\newcommand{\llangle}[1][]{\savebox{\@brx}{\(\m@th{#1\langle}\)}%
	\mathopen{\copy\@brx\mkern2mu\kern-0.9\wd\@brx\usebox{\@brx}}}
\newcommand{\rrangle}[1][]{\savebox{\@brx}{\(\m@th{#1\rangle}\)}%
	\mathclose{\copy\@brx\mkern2mu\kern-0.9\wd\@brx\usebox{\@brx}}}
\makeatother

\makeatletter
\newcommand*{\coloneqq}{\mathrel{\rlap{%
			\raisebox{0.28ex}{$\m@th\cdot$}}%
		\raisebox{-0.28ex}{$\m@th\cdot$}}%
	=}
\newcommand*{\eqqcolon}{=\mathrel{\rlap{%
			\raisebox{0.28ex}{$\m@th\cdot$}}%
		\raisebox{-0.28ex}{$\m@th\cdot$}}%
}
\makeatother

\makeatletter
\newcommand*{\rom}[1]{\expandafter\@slowromancap\romannumeral #1@}
\makeatother

\DeclareSymbolFont{cmbrightop}{OT1}{cmbr}{m}{n}
\DeclareMathSymbol{\sfPsi}{\mathalpha}{cmbrightop}{9}

\newcommand{\eqgraph}[3]{%
	\begin{gathered}
		\raisebox{0pt}[\dimexpr\height+#1][\dimexpr\depth+#2]{\ignorespaces#3\unskip}%
	\end{gathered}
}

\let\dim\relax
\newcommand{\dim}[1]{[\![ #1 ]\!]}

\allowdisplaybreaks

\begin{document}
	
	\title{Topology-driven deconfined quantum criticality in magnetic bilayers}
	
	\author{Sondre Duna Lundemo} 
	\affiliation{Center for Quantum Spintronics, Department of Physics, Norwegian University of Science and Technology, NO-7491 Trondheim, Norway}
	
	\author{Flavio S. Nogueira}
	\affiliation{Institute for Theoretical Solid State Physics, IFW Dresden, Helmholtzstr. 20, 01069 Dresden, Germany}
	
	\author{Asle Sudb\o}
	\email[Corresponding author: ]{asle.sudbo@ntnu.no}
	\affiliation{Center for Quantum Spintronics, Department of Physics, Norwegian University of Science and Technology, NO-7491 Trondheim, Norway}
	
	\date{\today} 
	
	\begin{abstract}
        Two-dimensional quantum antiferromagnets are believed to host phases of matter whose excitations are more fundamental than those of the ordered phases.
		When combining two such spin systems in a bilayer, strong interaction between the emergent excitations can produce phases not realized in either of its subsystems. 
		We show that the critical fluctuations of a two-dimensional spin liquid state can induce a deconfined quantum critical point in a proximate antiferromagnet. 
		The most relevant coupling between the associated effective field theories is given by a mixed Chern-Simons term of the emergent gauge fields in each layer. 
		This describes a topological current-current interaction. 
		In contrast to the local spin-spin interaction, it strongly modifies the renormalization group flow of the theory describing the Néel--valence-bond-solid transition of the antiferromagnet. 
		In particular, the protected coupling constant associated with it implies 
        non-trivial quantum critical scaling characterized by a non-universal power-law divergence of the correlation length in the critical domain and Berezinskii--Kosterlitz--Thouless divergence approaching it.
        
	\end{abstract}
	
	\maketitle 
	
	\section{Introduction}\label{sec:intro}
	
	Fractionalization is the remarkable phenomenon in which the low-energy excitations of a system appear to carry a fraction of the quantum numbers associated with the microscopic degrees of freedom \cite{chenSymmetryFractionalizationTwo2017}.
	The most prominent example of such an emergent phenomenon, originating from strong correlations, is the fractional quantum hall effect, discovered in a two-dimensional electron gas close to zero temperature when subjected to an extreme, out-of-plane magnetic field \cite{tsuiTwoDimensionalMagnetotransportExtreme1982,laughlinAnomalousQuantumHall1983,haldaneFractionalQuantizationHall1983}.
	Quantum spin liquids (QSLs) on the other hand are exotic, fractionalized states of spin systems completely bereft of classical order, but conjectured to be realizable under far less extreme experimental conditions \cite{zhouQuantumSpinLiquid2017}.
	The most accessible description of these states relies on the $\mathrm{SU}(2)$ projective construction, in which one rewrites the spin degree of freedom in terms of fractionalized degrees of freedom which are charged under an emergent $\mathrm{SU}(2)$ gauge field \cite{wenQuantumFieldTheory2007}.
	The true physical content of this description is revealed only after integrating out high-energy modes and assessing whether the fractionalized excitations are deconfined in the presence of the resulting low-energy effective gauge theory.
	Although demonstrating deconfinement is a formidable theoretical challenge, it is an essential one on the path to understanding the potential realizability of QSL states.
	The observation of such a state of matter has been a primary focus of experimental efforts in the field of quantum magnetism \cite{coldeaExperimentalRealization2D2001,wenExperimentalIdentificationQuantum2019,takagiConceptRealizationKitaev2019}.
	
	In recent years, a promising platform for the realization of QSL states has emerged in spin--orbit-coupled Mott insulators \cite{witczak-krempaCorrelatedQuantumPhenomena2014,rauSpinOrbitPhysicsGiving2016,jackeliMottInsulatorsStrong2009}, where strong correlations lead to the emergence of local magnetic moments with highly frustrated interactions originating with spin-orbit coupling.  
	An important subclass of these materials is so-called Kitaev materials, in which the localized spins interact through link-dependent Ising interactions as described by the Kitaev Hamiltonian or the related compass interactions \cite{kitaevAnyonsExactlySolved2006,nussinovCompassModelsTheory2015}.
	Competition between these interactions and the canonical Heisenberg exchange may promote quantum phase transitions (QPTs) between fractionalized and  conventionally ordered phases \cite{youDopingSpinorbitMott2012,shankarSymmetrybreakingEffectsInstantons2021,shankarContinuousTransitionIsing2022,schafferQuantumPhaseTransition2012,leeHeisenbergKitaevModelHyperhoneycomb2014,rauGenericSpinModel2014,saheliMajoranafermionMeanField2024,huNatureTopologicalPhase2024}.
	Moreover, bilayer systems of such materials have been conjectured to host exotic phases as a result of the interplay between fractionalized and electronic degrees of freedom \cite{senthilFractionalizedFermiLiquids2003,seifertFractionalizedFermiLiquids2018c,seifertBilayerKitaevModels2018,choiTopologicalSuperconductivityKondoKitaev2018,seifertFractionalizedFermionicQuantum2020,colemanSolvable3DKondo2022,tsvelikOrderFractionalizationKitaevKondo2022,luoTwistedBilayerU12022,lundemoTopologicalSuperconductivityInduced2024a,seifertWegnersIsingGauge2024,hwangAnyonCondensationConfinement2024}.
	Realistic models for spin--orbit-coupled Mott insulators therefore provide a natural laboratory for studying critical phenomena beyond the Landau-Ginzburg-Wilson (LGW) paradigm \cite{vojtaFrustrationQuantumCriticality2018,boyackQuantumPhaseTransitions2021}.  
	
	One of the most well-studied scenarios that evade the LGW paradigm is the quantum critical point separating ordered states of low-dimensional quantum systems where new elementary excitations not present in either of the ordered phases are liberated.
	These critical points are referred to as \textit{deconfined quantum critical points} (DQCPs) \cite{senthilDeconfinedQuantumCritical2004,senthilQuantumCriticalityLandauGinzburgWilson2004a,senthilDeconfinedCriticalityCritically2005}.
	Similar to a QSL state a DQCP describes a state of matter with fractionalized or deconfined excitations charged under an emergent gauge field.
	
\begin{figure}[htb]
		\centering
		\includegraphics[width=0.9\columnwidth]{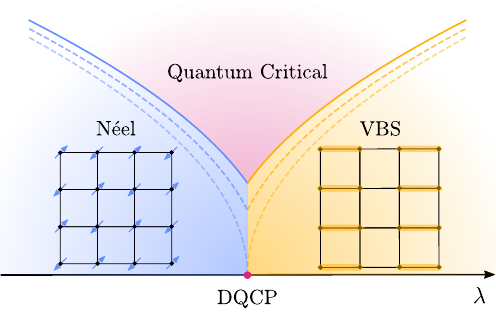}
		\caption{Schematic phase diagram of the Néel-VBS transition, as a function of some tuning parameter $\lambda$. The filled lines represent the order parameters of the two states and the associated preemptive, first-order transition that would follow from the LGW description. The dashed lines move the transition toward the DQCP. In our model, the parameter $\kappa$ interpolates between these scenarios.}
		\label{fig:pd}
	\end{figure}

	The canonical example of such a transition is the point separating the Néel state from the valence-bond solid (VBS) state of $S=1/2$ quantum antiferromagnets. 
	Based on the arguments of Refs.~\cite{senthilDeconfinedQuantumCritical2004,senthilQuantumCriticalityLandauGinzburgWilson2004a,senthilDeconfinedCriticalityCritically2005}, the universal aspects of this transition are governed by the noncompact $\mathbb{C}\mathrm{P}^{1}$ model with a Maxwell term in $2+1$ spacetime dimensions, also known as the Abelian Higgs model (AHM) 
	\begin{equation}\label{eq:LAHM}
		\calL_{\mathrm{AHM}} = \abs{\mathrm{D}_{\mu} z}^2 + m^2 \abs{z}^2 + \frac{u}{2}\abs{z}^4+ \frac{1}{2e^2} \left( \epsilon_{\mu\nu\rho} \partial_{\nu} A_{\rho} \right)^2,
	\end{equation}
	where  $\mathrm{D}_{\mu} \coloneqq \partial_{\mu} + \iu A_{\mu}$ denotes the covariant derivative.
	The doublet of complex fields appearing in this theory $z_{\alpha}$ ($\alpha=1,2$) is related to the fluctuations of sublattice magnetization in the quantum antiferromagnet through the Hopf map $\v{n} = \bar{z}_{\alpha} \bm{\sigma}_{\alpha\beta} z_{\beta}$ \cite{polyakovGaugeFieldsStrings1987}, and physically represent the spinons.
	The gauge field appearing in this effective field theory emerges from the continuum limit of a lattice theory and is therefore necessarily compact at the outset.
	Whenever the vacuum expectation value of $\abs{z}^2$ is zero, the compact nature of the gauge field necessitates the introduction of so-called \textit{plastic gauge fields} with support on the surfaces across which the emergent gauge field jumps by $2\pi$ \cite{kleinertMultivaluedFieldsCondensed2008,kleinertDeconfinementTransitionThreeDimensional2002a}.
	These excitations gap the dual gauge field and produce the VBS state.
	On the other hand, when $\abs{z}^2$ receives a non-zero vacuum expectation value the gauge field is massive due to the Higgs mechanism, giving rise to the Néel state. 
	The most fundamental conjecture of the deconfined quantum criticality (DQC) scenario is that the Higgs mass and the dual gauge-field mass vanish continuously as the transition is approached from each side respectively, leading to the suppression of instantons otherwise responsible for spinon confinement \cite{nogueiraDeconfinedQuantumCriticality2013}.   
    A schematic phase diagram for this transition is shown in Fig.~\ref{fig:pd}.
	In the present work, we therefore take the lagrangian in Eq.~\eqref{eq:LAHM} with a non-compact gauge field as a starting point, assuming it describes the DQCP.
	
	The proposed effective field theory describing the DQCP takes the form of the Ginzburg-Landau theory for superconductivity, with the only difference being that the $z$ field has two complex components and the superconducting order parameter has one. 
	One would therefore expect the Néel-VBS transition to display critical behavior (in terms of the $z$ bosons) similar to that of the superconducting transition. 
	However, a subtle point regarding the AHM is that standard field-theoretical methods applied to it predict the phase transition to be first-order \cite{nogueiraFieldTheoreticApproaches2004b}.
	In particular, a one-loop renormalization group (RG) analysis of this theory in $d=4-\epsilon$ spacetime dimensions predicts an infrared-stable charged fixed point only if the number of complex components of the matter field is artificially extended to $N_b\geq 183$ \cite{halperinFirstOrderPhaseTransitions1974,colemanRadiativeCorrectionsOrigin1973}.
	Higher-order loop expansions show a dramatic decrease in the critical $N_b$ \cite{kolnbergerCriticalFluctuationsSuperconductors1990,ihrigAbelianHiggsModel2019}, but require the use of sophisticated resummation techniques for which it is challenging to provide a convincing rationale.
	On the other hand, duality arguments point to the existence of a charged critical point even for $N_b=1$ in the type-\rom{2} regime. \cite{dasguptaPhaseTransitionLattice1981,kiometzisCriticalExponentsSuperconducting1994}.
	
	The very large $N_b$ required to obtain a critical point of the theory is a symptom of the perturbative RG's inability to correctly capture the non-perturbative effects of singular fluctuations in the phase of the complex field \cite{kleinertMultivaluedFieldsCondensed2008}.
	In fact, the relevance of such fluctuations is controlled in this model by the ratio of the two mass scales that emerge at the superconducting transition: the correlation length and the magnetic penetration depth \cite{kleinertDisorderVersionAbelian1982,herbutCriticalFluctuationsSuperconductors1996,moOrderMetaltosuperconductorTransition2002}.
	Aside from making the RG treatment more tractable, the introduction of the number of complex components $N_b$ therefore provides a theoretical knob tuning between a weakly first-order and second-order transition, which represent the critical behavior for type-\rom{1} and type-\rom{2} superconductors respectively.
	However, in the context of two-dimensional quantum antiferromagnetic ($N_b=2$) the case remains more elusive, with a large body of numerical work in favor of a weakly first-order transition instead of the DQC scenario  \cite{kuklovDeconfinedCriticalityRunaway2006,kragsetFirstOrderPhaseTransition2006,kuklovDeconfinedCriticalityGeneric2008,herlandPhaseTransitionsThree2010,herlandPhaseStructurePhase2013,zhangContinuousEasyPlaneDeconfined2018,desaiFirstorderPhaseTransitions2020,bonatiLatticeAbelianHiggsModel2021,bonatiAbelianHiggsGauge2023,bonatiDiverseUniversalityClasses2024}.
	As such, we can interpret the large $N_b$ required to obtain a critical point in the spinon theory described by Eq.~\eqref{eq:LAHM} as a deviation from the DQC scenario, and implying instead \textit{deconfined pseudocriticality} \cite{maTheoryDeconfinedPseudocriticality2020}.
	After taking this point of view, the search for a realization of the DQCP immediately leads us to contemplate what can reduce the critical $N_b$. 

	Motivated by this question and the recent theoretical progress on the subject of bilayer QSLs \cite{luoTwistedBilayerU12022,hwangAnyonCondensationConfinement2024}, we study an extension of the AHM that describes the low-energy effective field theory of a {\it bilayer of quantum antiferromagnets}, where one is supposed to realize a $\mathrm{U}(1)$ Dirac spin liquid and the other is close to the Néel-VBS transition.
	By the arguments above, the latter has a description in terms of the AHM. 
	The former is described by quantum electrodynamics in $d=3$ spacetime dimensions with $N_f=4$ flavors of massless Dirac fermions.
	We suggest a coupling between these theories given by a mixed Chern-Simons term, also known as a ``background-field" (BF) term.
	In contrast to local matter interactions, this represents an interaction between matter fields mediated by the emergent gauge fields and is thus unique to fractionalized phases. 
	Hence, the very presence of these gauge fields may give rise to an interaction  
    with the potential to promote criticality in the quantum antiferromagnet.
	
	In this paper, we perform an RG analysis of this theory in $d=3$ spacetime dimensions. 
	By analyzing the one-loop RG flow, we compute the critical number of bosons $N_b$ required to obtain a second-order phase transition, with the aim of elucidating the effect of the fractionalized excitations of the QSL on the DQCP.
	Moreover, we compute the correlation length in the critical regime and the anomalous dimension $\eta$ of the two-spin correlation function, defined by
	$
		\mathcal{G}(x) \coloneqq \left\langle \v{n}(x) \cdot \v{n}(0) \right\rangle \sim 1/\abs{x}^{d-2+\eta},
	$
	in the antiferromagnet.
	Lastly, we study the system's predisposition to chiral symmetry breaking to understand whether the coupling between the sublayers can be detrimental to the stability of the QSL state.
	
	\section{Effective model}\label{sec:Model}
	
	We describe the bilayer system with the Euclidean Lagrangian
	\begin{subequations}\label{eq:L}
		\begin{align}
			\calL &\coloneqq \calL_{\mathrm{AHM}} + \calL_{\mathrm{DSL}} + \calL_{\mathrm{BF}},
			\intertext{where $\calL_{\mathrm{AHM}}$ is given by Eq.~\eqref{eq:LAHM} and the Dirac spin liquid is described with}
			\calL_{\mathrm{DSL}} &\coloneqq \bm{\bar{\psi}} \cdot \left( \slashed{\partial} + \iu \slashed{a} \right) \bm{\psi} + \frac{1}{2f^2} \left( \epsilon_{\mu\nu\rho} \partial_{\nu} a_{\rho} \right)^2,
		\end{align}
		where $\bar{\psi}_{a}, \psi_{a} $ with $a=1,\dotsb , N_f$ are four-component Grassmann-valued fields. Here $\slashed{X} \coloneqq \gamma_{\mu} X_{\mu}$ denotes the contraction between a vector $X_{\mu}$ and the $4\times4$ Dirac matrices $\gamma_{\mu}$ obeying the Clifford algebra $\{ \gamma_{\mu}, \gamma_{\nu} \} = 2 \delta_{\mu\nu}$.
		The most natural way two spin systems couple in a bilayer is through a local spin-spin interaction ($s$-$d$-type).
		However, when the spin excitations are fractionalized into bosonic spinons in one layer and fermionic in the other, such a coupling has scaling dimension $d-3$ and is therefore perturbatively irrelevant in $d=4$, which is the upper critical dimension of the AHM. 
		This interaction is therefore irrelevant as far as the criticality of the Néel-VBS transition is concerned (see Appendix~\ref{app:biquadratic} for a more elaborate justification).
		
		Despite rendering the local matter coupling irrelevant, the fractionalization of the spin offers a fundamentally different way for the sublayers to couple. 
		In $d=3$ the two emergent $\mathrm{U}(1)$ gauge fields can couple via a topological BF term
		\begin{equation}
			\calL_{\mathrm{BF}} \coloneqq \iu \frac{\kappa}{4\pi} \epsilon_{\mu\nu\rho} a_{\mu} \partial_{\nu} A_{\rho}.
		\end{equation}
	\end{subequations}
    This is, in fact, the most relevant, gauge-invariant, local interaction between $\calL_{\mathrm{AHM}}$ and $\calL_{\mathrm{DSL}}$ one can consider.
	The motivation for our positing such a term is twofold. 
	First, it is intriguing to study a coupling between spin systems that can appear alongside fractionalization and is absent in the symmetry-broken regimes.
	Additionally, we can argue that it has a natural microscopic interpretation in terms of spinon currents.
	  Prior to resolving the constraint softly and including the Maxwell term in the $\mathbb{C}P^{1}$ model to arrive at the paradigmatic model in Eq.~\eqref{eq:LAHM}, the $A_{\mu}$ gauge field is a non-dynamical auxiliary field \cite{sachdevEffectiveLatticeModels1990,sachdevGroundStatesQuantum2002}. 
	Its equation of motion identifies it with the spinon current
    \footnote{Note that Eq.~\eqref{eq:Aaux} does not follow directly from Eq.~\eqref{eq:LAHM}. 
    Instead, it is the equation of motion for the $A_{\mu}$ field when using the $\mathbb{C}P^{N_b-1}$ model
    \begin{equation}\label{eq:cp1}
        S = \frac{1}{2g} \int \mathrm{d}^d x \abs{(\partial_{\mu} + \mathrm{i} A_{\mu}) z}^2,
    \end{equation}
    when the $z$ bosons are subjected to the hard constraint $\abs{z}^2 = 1$. 
    The $\mathbb{C}P^{N_b-1}$ model is the correct starting point for discussing the Néel-VBS transition. However, as mentioned in the Introduction, due to a subtle cancellation of instantons and Berry phases \cite{senthilDeconfinedQuantumCritical2004,kragsetFirstOrderPhaseTransition2006}, the universality of this transition is governed by the non-compact $\mathbb{C}P^{N_b-1}$ model. 
    }
\begin{equation}\label{eq:Aaux}
		A_{\mu} = \frac{\iu}{2} \left[  z^{\dagger} \partial_{\mu} z - \left(\partial_{\mu} z^{\dagger} \right) z  \right].
	\end{equation}  
    Moreover, using Eq. \eqref{eq:Aaux} and the Hopf map, it can be shown that the field strength is related to the density of topological charge (skyrmion density) as
	$
		F_{\mu\nu} = \v{n} \cdot \left(\partial_{\mu} \v{n} \times \partial_{\nu} \v{n} \right)/2
	$ \cite{polyakovGaugeFieldsStrings1987}.
	The analogous origin of the emergent gauge field $a_{\mu}$ of the Dirac spin liquid permits interpreting the BF term as a coupling between the spinon current of one layer and the density of topological charge in the other.
    A partial integration interchanges the roles of $a_{\mu}$ and $A_{\mu}$, reflecting a duality in this description \cite{hanssonSuperconductorsAreTopologically2004}. 
    
    The topological BF term has previously been considered in other effective theories of bilayers of fractionalized systems \cite{wenTunnelingDoublelayeredQuantum1993,zouDeconfinedMetalinsulatorTransitions2020,sodemannCompositeFermionDuality2017}.
    Based on the irrelevance of the local spin-spin interaction demonstrated in Appendix~\ref{app:biquadratic} it is conceivable that the BF term can derive from a non-local spin-spin interaction. This would be analogous to its origin in theories for quantum-Hall bilayers, which is through inter-layer Coulomb interaction.

   \section{Renormalization group analysis}\label{sec:RG}
	
	Let us derive the RG flow of the theory defined by Eq.~\eqref{eq:L} in fixed dimension $d=3$, employing a regularization scheme in which all diagrams are computed with a non-zero external momentum transfer that defines the renormalization scale $\mu^2 \coloneqq q^2$ \cite{herbutCriticalFluctuationsSuperconductors1996,nogueiraFieldTheoreticApproaches2004b,kleinertCriticalProperties$phi^4$2001}.
	Following the seminal work of Ref.~\cite{halperinFirstOrderPhaseTransitions1974}
    on the critical properties of the AHM, we generalize the $z$ field to an $N_b$-component complex boson, keeping in mind that the physical value is $N_b=2$.
	We also leave the number of fermion flavors $N_f$ unspecified.
	To facilitate the RG procedure, we demote the coupling constants and fields in the Lagrangian to bare ones, denoted by a subscript $0$, and introduce renormalization constants which define the corresponding renormalized quantities according to
	$\phi_{0} = Z_{\phi}^{1/2} \phi$ for the fields and $\lambda_ {0} = Z_{\lambda} Z_{\mathcal{O}}^{-1} \lambda$ for the coupling constant $\lambda$ associated to the operator $\mathcal{O}$.
	With the help of the renormalization scale $\mu$ we also define the dimensionless coupling constants as $\hat{m} \coloneqq m / \mu, \hat{u} \coloneqq u/\mu^{4-d}, \hat{e}^2 \coloneqq e^2/\mu^{4-d}$, and $\hat{f}^2 \coloneqq f^2/\mu^{4-d}$.
	For convenience, we rescale the gauge fields $A_{\mu} \mapsto e A_{\mu} $ and $a_{\mu} \mapsto f a_{\mu}$ and define the quantity $g \coloneqq \kappa e f /4\pi$.
	
	Imposing gauge invariance implies the constraints $Z_{z} = Z_{e} = Z_{e^2}$ and $Z_{\psi} = Z_{f}$.
	The renormalized Lagrangian takes the form
	\begin{equation}\label{eq:LR}
		\begin{split}
			\mathcal{L} &= Z_{z} \abs{\left(\partial_{\mu} + \iu \hat{e} \mu^{2-d/2} A_{\mu}\right) \v{z}}^2 + Z_{2} \hat{m}^2 \mu^2 \abs{\v{z}}^2 \\
			&+ Z_{u} \frac{\hat{u}}{2} \mu^{4-d} \left(\abs{\v{z}}^2\right)^2 +Z_{\psi} \bar{\bm{\psi}} \cdot \left( \slashed{\partial} + \iu \hat{f} \mu^{2-d/2} \slashed{a}\right) \bm{\psi} \\[0.4em]
			&+ Z_{A} \frac{1}{4} F_{\mu\nu}^2 + Z_{a} \frac{1}{4} f_{\mu\nu}^2 + Z_{g} \iu \hat{g} \mu \epsilon_{\nu\rho\lambda}  a_{\nu} \partial_{\rho} A_{\lambda}. 
		\end{split}
	\end{equation}
	Since the BF term is a topological term it does not depend on the metric and hence $\kappa$ does not scale \cite{colemanNoMoreCorrections1985}, and therefore $Z_{g} = 1/\sqrt{Z_{a} Z_{A}}$.
	 
	\begin{fmffile}{diagrams}  
		\fmfset{dash_len}{1.5mm}
		\fmfset{wiggly_len}{2mm}
		\fmfset{dot_size}{3pt}
		
		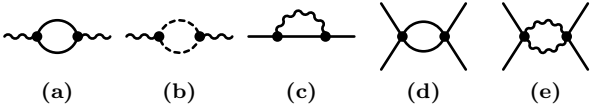
\begin{figure}[htb]
			\centering
			\captionsetup[subfloat]{labelfont=bf}
			\subfloat[\label{subfig:a}]{
				\centering
				\begin{fmfgraph*}(40,25)
					\fmfkeep{ZA}
					\fmfleft{i}
					\fmfright{r}
					\fmf{boson,tension=1}{i,v1}
					\fmf{plain,tension=0.4,right=0.8}{v1,v2}
					\fmf{plain,tension=0.4,right=0.8}{v2,v1}
					\fmf{boson,tension=1}{v2,r}
					\fmfdot{v1,v2}
				\end{fmfgraph*}
			}
			\subfloat[\label{subfig:b}]{
				\centering
				\begin{fmfgraph*}(40,25)
					\fmfkeep{Za}
					\fmfleft{i}
					\fmfright{r}
					\fmf{boson,tension=1}{i,v1}
					\fmf{dashes,tension=0.4,right=0.8}{v1,v2}
					\fmf{dashes,tension=0.4,right=0.8}{v2,v1}
					\fmf{boson,tension=1}{v2,r}
					\fmfdot{v1,v2}
				\end{fmfgraph*}
			}
			\subfloat[\label{subfig:c}]{
				\centering
				\begin{fmfgraph*}(40,25)
					\fmfkeep{Zz}
					\fmfleft{i}
					\fmfright{r}
					\fmf{plain}{i,v1}
					\fmf{plain,tension=0.3}{v1,v2}
					\fmf{plain}{v2,r}
					\fmf{boson,left=1,tension=0.3}{v1,v2}
					\fmfdot{v1,v2}
				\end{fmfgraph*}
			}
			\subfloat[\label{subfig:d}]{
				\centering
				\begin{fmfgraph*}(40,25)
					\fmfkeep{Zu1}
					\fmfleft{l1,l2}
					\fmfright{r1,r2}
					\fmf{plain}{l1,v1,l2}
					\fmf{plain,right=0.7,tension=0.5}{v1,v2,v1}
					\fmf{plain}{r1,v2,r2}
					\fmfdot{v1,v2}
				\end{fmfgraph*}
			}
			\subfloat[\label{subfig:e}]{
				\centering
				\begin{fmfgraph*}(40,25)
					\fmfkeep{Zu2}
					\fmfleft{l1,l2}
					\fmfright{r1,r2}
					\fmf{plain}{l1,v1,l2}
					\fmf{boson,right=0.7,tension=0.5}{v1,v2,v1}
					\fmf{plain}{r1,v2,r2}
					\fmfdot{v1,v2}
				\end{fmfgraph*}
			}
			\caption{Non-vanishing, one-particle-irreducible, one-loop diagrams contributing to the renormalization of $e^2$ \textbf{(a)}, $f^2$ \textbf{(b)} and $u$ \textbf{(c)}-\textbf{(e)} in the effective theory. The plain lines represent the fluctuating $z$ field, the wiggly line the $\mathrm{U}(1)$ gauge field $\mathcal{A}_{\mu} \coloneqq \left( A_{\mu} \, a_{\mu} \right)^{\mathsf{T}}$ and the dashed lines the Dirac fermion $\psi$.
            The vertices involving the gauge boson implicitly carry an index specifying whether the matter fields couple to $A_{\mu}$ or $a_{\mu}$.
            }
			\label{fig:feynman_diagrams_1}
		\end{figure}
	
	\end{fmffile}

    \subsection{Beta functions}\label{sec:beta_functions}

    \begin{figure*}[htb]
		\centering
		\includegraphics[width=\textwidth]{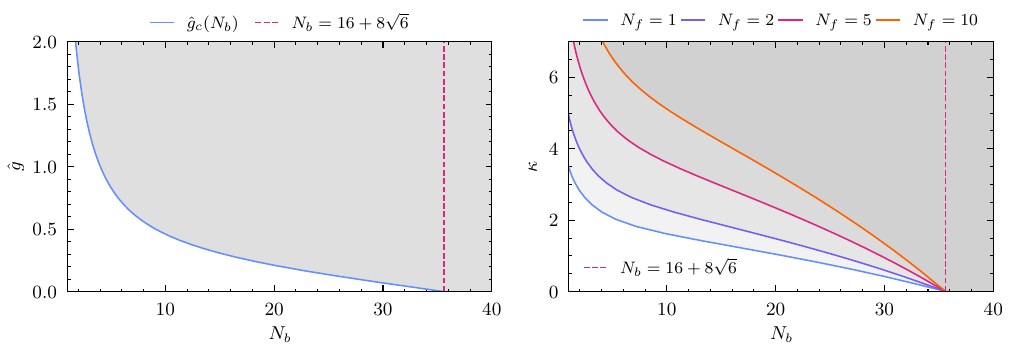}
		\caption{Region of existence of charged fixed point. The charged fixed point of the theory exists in the parameter regime shaded in grey, which is separated from the trivial phase (in white) by the filled lines. 
        The left-hand panel shows the critical line $\hat{g}_{c}(N_b)$, while the right-hand panel shows the critical lines $\kappa_{c}(N_b;N_f)$ for a selection of values of $N_f$.}
		\label{fig:fixed_pt_existence}
	\end{figure*}
    
    Focusing on the critical plane defined by $m^2 = 0$ and fixing the Landau gauge, the non-vanishing relevant diagrams reduce to those listed in Fig.~\ref{fig:feynman_diagrams_1}.
	These diagrams are computed using the massless limit of renormalized Lagrangian in Eq.~\eqref{eq:LR} and suffice to determine the renormalization constants of the theory.
	The $\beta$ function for the dimensionless coupling constant $\hat{\lambda} $ ($=\hat{e}^2, \hat{f}^2, \hat{u}$) is defined as $\beta_{\lambda} \coloneqq \partial \hat{\lambda}/\partial \log \mu$, where the derivative is evaluated at fixed bare couplings.
	The $\beta$ functions for $e^2$, $f^2 $ and $u$ are given by (See Appendix \ref{app:RG} for details) 
	\begin{widetext}
	\begin{subequations}\label{eq:beta_functions_d}
		\begin{align}
			\beta_{e^2} &= (d-4) \left[ \hat{e}^2 + 2 N_b F(d) \left( \hat{e}^2 \right)^2 \right] \\
			\beta_{f^2} &= (d-4) \left[ \hat{f}^2 + 4 N_f (d-2)F(d) \left( \hat{f}^2 \right)^2 \right] \\
			\beta_{u} &= (d-4) \left[ \hat{u} - (N_b + 4) G(d) \hat{u}^2 - 2 \left(\hat{e}^2\right)^2 \mathcal{B}(\hat{g};d) + 2 \hat{e}^2 \hat{u} \mathcal{W}(\hat{g};d) \right] - 2 \left(\hat{e}^2\right)^2 \mu \frac{\partial}{\partial \mu}  \mathcal{B}(\hat{g};d) + 2 \hat{e}^2 \hat{u} \mu \frac{\partial}{\partial \mu} \mathcal{W}(\hat{g};d),
		\end{align}
	\end{subequations}
	where the functions $F(d)$ and $G(d)$ are defined by
	\begin{equation}\label{eq:F&G}
		F(d) \coloneqq \frac{1}{(4\pi)^{d/2}} \Gamma\left(1 - \frac{d}{2} \right)  \frac{\Gamma^2(d/2)}{\Gamma(d)} \quad \text{and} \quad G(d) \coloneqq \frac{1}{(4\pi)^{d/2}} \Gamma\left(2 - \frac{d}{2} \right) \frac{\Gamma^2(d/2-1)}{\Gamma(d-2)},
	\end{equation}
	while $\mathcal{B}(\hat{g})$ and $\mathcal{W}(\hat{g})$ are defined as the $\hat{g}$ dependence of the counterterms corresponding to the bubble and wave diagram in Figures \ref{subfig:e} and \ref{subfig:c} respectively.
    Recalling that $\kappa$ does not flow, the flow of $g= \kappa e f /(4\pi)$ is determined exclusively by that of $e^2$ and $f^2$, viz.,
    \begin{equation}\label{Eq:beta-g}
        \beta_{g}=\frac{1}{2}\left(\frac{\beta_{e^2}}{\hat{e}^2}+\frac{\beta_{f^2}}{\hat{f}^2}\right)\hat{g}.
    \end{equation}
    Thus, at the fixed points $\hat{e}_*^2$ and $\hat{f}_*^2$ it follows that $\beta_g$ vanishes for any $\hat{g}_*=\kappa\hat{e}_*\hat{f}_*/(4\pi)$. 
    This leads to a scaling behavior parametrized by $\kappa$.  

	In $d=3$, we find that the $\beta$ functions in Eq.~\eqref{eq:beta_functions_d} simplify to
	\begin{subequations}\label{eq:beta_functions_3}
		\begin{align}
			\beta_{e^2} &= - \hat{e}^2 + \frac{N_b}{16} \left( \hat{e}^2 \right)^2  \label{eq:betae2} \\
			\beta_{f^2} &= - \hat{f}^2 + \frac{N_f}{8} \left( \hat{f}^2 \right)^2 \label{eq:betaf2} \\
			\beta_{u} &= - \hat{u} + \frac{N_b +4}{8} \hat{u}^2 
			+ 2 \hat{e}^4 \frac{\partial }{\partial \hat{g}} \left[ \hat{g} \mathcal{B}(\hat{g}) \right]-  2 \hat{e}^2 \hat{u} \frac{\partial }{\partial \hat{g}} \left[ \hat{g} \mathcal{W}(\hat{g}) \right], \label{eq:betau}
		\end{align}
	\end{subequations}
	where
	\begin{align}\label{eq:B}
		\mathcal{B}(\hat{g}) &= \frac{1}{32 \hat{g}^4} + \frac{1}{8\pi \abs{\hat{g}}} 
		- \left( \frac{1}{2} + \frac{1}{\hat{g}^2} + \frac{1}{2\hat{g}^4} \right) \frac{1}{4\pi} \arctan\left(\frac{1}{\abs{\hat{g}}}\right) +\left( 2 + \frac{1}{\hat{g}^2} + \frac{1}{4\hat{g}^2} \right)\frac{1}{4\pi} \arctan\left(\frac{1}{2\abs{\hat{g}}}\right), 
	\end{align}
	and
	\begin{equation}\label{eq:W}
			\mathcal{W}(\hat{g}) = \left[ 4 + \frac{1}{\hat{g}^2} \left(\hat{g}^2 - 1\right)^2 \right] \frac{1}{4\pi}\arctan\left( \frac{1}{\abs{\hat{g}}}\right) - \frac{1}{8\hat{g}^2} - \left(\hat{g}^2 - 1\right)\frac{1}{4 \pi \abs{\hat{g}}}. 
	\end{equation}
    \end{widetext}
    
    In the decoupled limit $\hat{g}\to 0$, we find 
    \begin{equation}\label{eq:decoupled}
	   \lim_{\hat{g} \to 0} \mathcal{B}(\hat{g}) = \frac{3}{16} \quad \text{and} \quad \lim_{\hat{g} \to 0} \mathcal{W}(\hat{g}) = \frac{1}{4}.
    \end{equation}
    so that 
    \begin{equation}
	   \beta_{u}(\hat{g}=0) = - \hat{u} + \frac{N_b+4}{8} \hat{u}^2 + \frac{3}{8} \hat{e}^4 - \frac{1}{2} \hat{e}^2 \hat{u}.
    \end{equation}
    This describes the criticality of the AHM.
    Moreover, in the asymptotic limit $\hat{g}\to \infty$
    \begin{equation}\label{eq:inf_limit}
	   \mathcal{B}(\hat{g}) \sim \frac{1}{4\pi \hat{g}} \quad \text{and} \quad \mathcal{W}(\hat{g}) \sim \frac{2}{3\pi \hat{g}}.
    \end{equation}
    From Eq.~\eqref{eq:beta_functions_3} we see that this leads to decoupled $\beta$ functions for $e^2$, $f^2$ and $u$, and in particular that $\beta_{u}$ reduces to the $\beta$ function for a $\phi^4$ theory with global $\mathrm{U}(N_b)$ symmetry.
    Hence, at large $\hat{g}$, interactions mediated by the gauge fields are screened, leading to completely deconfined spinons. 
    
    \subsection{Charged fixed points}
    
    A DQCP corresponds to a critical point where both the $z$ bosons and the gauge field $A$ are well-defined excitations.
	In this language, it corresponds to an infrared-stable (IR-stable), non-trivial ($\hat{u}_{*}\neq0$) charged ($\hat{e}^2 \neq 0 $) fixed point. 
    From Eq.~\eqref{eq:betae2} it is clear that there is a charged fixed point at $\hat{e}^{2}_{*} = 16/N_b$, which is IR-stable irrespective of the other coupling constants. 
    The subsystems are only coupled if both $e_{*} \neq 0$ and $f_{*} \neq 0$, which implies that the charged fixed point of Eq.~\eqref{eq:betaf2} with $\hat{f}_{*}^2 = 8/N_f$ is the most interesting.

    The $\beta$ functions admit non-trivial charged fixed points whenever the discriminant of the second-order algebraic equation obtained by replacing $\hat{e}^2_{*} = 16/N_b$ in Eq.~\eqref{eq:betau} is non-negative.
    That is, $D(\hat{g},N_b) \geq 0$ where
    \begin{equation}
        \begin{split}
            D(\hat{g},N_b) &\coloneqq \left( 1 + \frac{32}{N_b} \frac{\partial}{\partial \hat{g}} \left[ \hat{g} \mathcal{W}(\hat{g}) \right] \right)^2 \\
            &\qquad\qquad\qquad- 512\frac{N_b + 4}{2N_b^2} \frac{\partial}{\partial \hat{g}} \left[ \hat{g} \mathcal{B}(\hat{g}) \right],
        \end{split}
    \end{equation}
    or, equivalently
    \begin{equation}
    	\left( N_b + 32 \frac{\partial}{\partial \hat{g}} \left[ \hat{g} \mathcal{W}(\hat{g}) \right] \right)^2 \geq 256\left(N_b + 4\right) \frac{\partial}{\partial \hat{g}} \left[ \hat{g} \mathcal{B}(\hat{g}) \right].
    \end{equation}
    This condition implicitly defines a line $\hat{g}_{c}(N_b)$ separating regimes of vastly different scaling.
    The region of existence of a charged fixed point in the $(\hat{g},N_b)$ plane as well as the $(\kappa, N_b)$ plane is illustrated in Fig.~\ref{fig:fixed_pt_existence}.
    The latter being inferred from substituting $\hat{g} = \kappa \times 2 \sqrt{2} / \pi \sqrt{N_f N_b}$ in the expression for the discriminant. 
    
    From Fig.~\ref{fig:fixed_pt_existence} it is clear that finite $\hat{g}$ causes a dramatic decrease in the critical number of bosons $N_b$ required for a charged fixed point.
    Analytically, this behavior can be deduced by comparing the decoupled limit in Eq.~\eqref{eq:decoupled} to the infinite-coupling limit in Eq.~\eqref{eq:inf_limit}, from which one realizes that the topological mass $\hat{g}$ interpolates between the AHM and the $N_b$-component complex $\phi^4$ theory.
    The latter of which displays a non-trivial fixed point irrespective of the value of $N_b$.
    Note that in the decoupled limit $\hat{g}\to 0$ we obtain the critical $N_{b} = 8\left(2 + \sqrt{6}\right) \simeq 35.6$ which is considerably lower than the corresponding value obtained using the $\epsilon$ expansion around $d=4$ \cite{halperinFirstOrderPhaseTransitions1974}.
    The reason for this improvement is that fixed-dimension approaches are more well-behaved for this problem \cite{herbutCriticalFluctuationsSuperconductors1996,kleinertChargedFixedPoint2003}.
    
    \subsection{Scaling and critical exponents}\label{sec:scaling}

    \begin{figure*}[t]
		
		\centering
		\raisebox{-0.11\height}{\includegraphics[width=.33\textwidth]{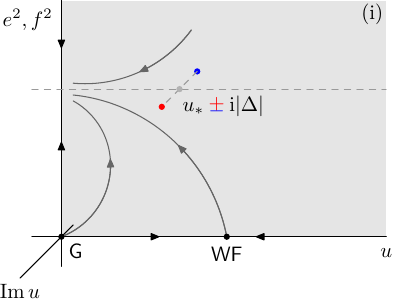}}\hfill
		\includegraphics[width=.33\textwidth]{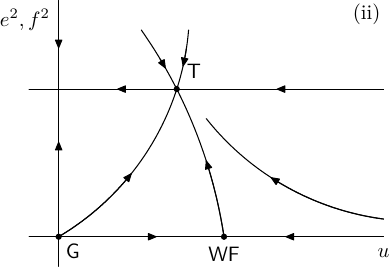}\hfill
		\includegraphics[width=.33\textwidth]{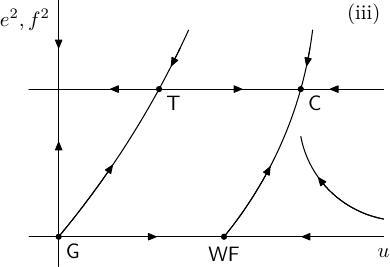}
		
		\caption{Schematic flow diagram in the three different regimes reachable by tuning $g$ (or equivalently, $\kappa$): (i) $D(\hat{g}) < 0$, (ii) $D(\hat{g}) = 0$ and (iii) $D(\hat{g}) > 0 $. Approaching the limiting scenario $D(\hat{g}) = 0$ yields different critical scaling dependent on whether $u > u_{*}$ or $u < u_{*}$. In this figure $\mathsf{G}$ denotes the Gaussian fixed point, $\mathsf{WF}$ the Wilson-Fisher fixed point, $\mathsf{T}$ the tri-critical point and $\mathsf{C}$ the charged fixed point.
		The illustration is inspired by Fig.~1 in Ref.~\cite{ihrigAbelianHiggsModel2019}.
		}
		\label{fig:schematic_flow}
		
	\end{figure*}

    Since the lines separating the regions where charged fixed points exist and where they do not exist are defined by the discriminant changing sign, the critical point changes nature when approaching this line \cite{kaplanConformalityLost2009}.
    In general, we can express the $\beta$ function for $u$ at the charged fixed point as 
    \begin{equation}
    	\beta(u;\hat{g}) \coloneqq \beta_{u} \bigr\lvert_{\hat{e}^2 = \hat{e}^2_{*}} = \frac{N_b + 4}{8} \left( \hat{u} - \hat{u}_{+} \right)\left( \hat{u} - \hat{u}_{-} \right), 
    \end{equation}
    where
    \begin{align}
    	\hat{u}_{\pm} &= \frac{4}{N_b + 4} \left\{ \left( 1 + \frac{32}{N_b} \frac{\partial}{\partial \hat{g}} \left[ \hat{g} \mathcal{W}(\hat{g}) \right] \right) \pm \sqrt{D(\hat{g},N_b)} \right\} \notag \\
    	&\equiv \hat{u}_{*} \pm \Delta,
    \end{align}
    where both $u_{*}$ and $\Delta$ depend implicitly on $\hat{g}$ and $N_b$ as above.
    Note that for $\sign{D} = 1$ we have $\Delta = \abs{\Delta}$, while for $\sign{D} = -1$ we have $\Delta = \iu \abs{\Delta}$.
    This lets us write the $\beta$ function compactly as 
    \begin{equation}
    		\beta(u;\hat{g}) = \frac{N_b + 4}{8} \left( (\hat{u}-\hat{u}_{*})^2 - \sign{D}(\hat{g},N_b) \abs{\Delta}^2 \right).
    \end{equation}
    The schematic flow described by the $\beta$ functions is illustrated in Fig.~\ref{fig:schematic_flow}.
	
	\subsubsection{Regime $D > 0$}
    The negated $\beta$ function describes the flow of $\hat{u}$ under scaling of lengths $x \mapsto x'(l) \coloneqq l^{-1} x$. 
    By integrating it, one can deduce the function $\xi'(l)$ required to leave the correlation length $\xi \equiv l \xi'(l)$ scale-invariant.
    In the critical regime $D(\hat{g},N_b) > 0$, integrating the $\beta$ function yields the scaling of the correlation length
    \begin{equation}
        \xi \sim \abs{\frac{\hat{u} - \hat{u}_{-}}{\hat{u} - \hat{u}_{+}}}^{4/(N_b+4) \abs{\Delta}},
    \end{equation}
    identifying $u_{+} \equiv u_{\mathrm{IR}}$ as the IR-stable fixed point.
    Note that $u_{-}$ would appear to be UV-stable from this equation, but is in fact a tri-critical point due to UV-unstable flow along the $\hat{e}^2$ and $\hat{f}^2$ directions.
    
    In the critical regime, we can compute the scaling of the two-spin correlation function 
    \begin{equation}
        \begin{split}
            \mathcal{G}(x) \equiv 2 \langle \bar{\v{z}}(x) \cdot \v{z}(0) \v{z}(x) &\cdot \bar{\v{z}}(0) \rangle  \\
            &- \frac{2}{N_b} \left\langle \abs{\v{z}(x)}^2 \abs{\v{z}(0)}^2 \right\rangle.
        \end{split}
    \end{equation}
    While the scaling behavior of the second term can be obtained by examining the scaling dimension of a mass term $\abs{\v{z}}^2$, the first term is associated with an anisotropic mass \cite{nogueiraQuantumCriticalScaling2007,nogueiraDeconfinedQuantumCriticality2008}.
    When extending $z$ to be an $N_b$-component complex scalar, we let $z_1$ and $z_2$ have $N_b/2$ complex components each, and consider anisotropic operator insertions of $\abs{z_1}^2$ and $\abs{z_2}^2$ in the correlation function $\langle \bar{z}_1(x)z_{1}(0)\rangle$.
    By noting that the scaling dimension of the mass term is given by
    $
        \dim{\abs{z_1}^2 + \abs{z_2}^2} = d - 1/\nu,
    $
    and that the scaling dimension of each component of $\v{n}$ is $d-\phi/\nu$, and particularly that $\dim{\abs{z_1}^2 - \abs{z_2}^2} = d - \phi/\nu$, we can obtain both $\nu$ and the crossover exponent $\phi$ by considering the sum and difference of insertions of $\abs{z_1}^2$ and $\abs{z_2}^2$. 
    These require renormalization constants $Z_2$ and $Z_2'$ respectively, which in turn define the anomalous dimensions 
    \begin{equation}\label{eq:eta2_def}
	   \eta_2 =  \mu \frac{\partial \log \left( Z_2/Z_z \right)}{\partial \mu} \biggr\lvert_{\mathrm{IR}} \,\, \text{\&} \,\, \eta_2' =  \mu \frac{\partial \log \left( Z_2'/Z_z \right)}{\partial \mu} \biggr\lvert_{\mathrm{IR}}
    \end{equation}
    where $\vert_{\mathrm{IR}}$ refers to evaluating the derivative at the IR-stable fixed point.
    
\begin{figure*}[t]
		
		\centering
		\includegraphics[width=\textwidth]{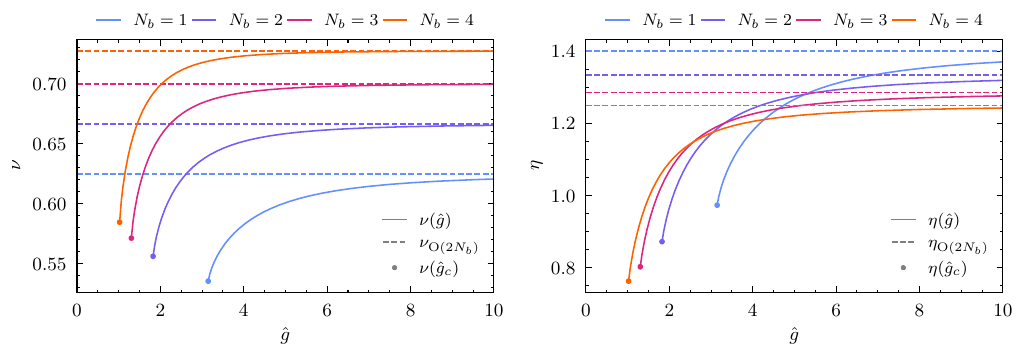}
		
		\caption{Critical exponents $\nu$ and $\eta$ as a function of the topological mass $\hat{g}$ for a selection of $N_b$ values. 
		These exponents asymptotically approach the ones associated with the $\mathrm{O}(2N_b)$ universality class shown in Eq.~\eqref{eq:O2N} as $\hat{g}\to\infty$.
		}
		\label{fig:critical_exponents}
	\end{figure*}

    The critical exponent $\nu^{-1} = 2 + \eta_2$ is found to be (See Appendix~\ref{app:critical_exponents} for details) 
    \begin{equation}\label{eq:nuinv}
        \nu^{-1}(\hat{g}) = 2 - \frac{N_b + 1}{8} \hat{u}_{\mathrm{IR}} + \hat{e}^2_{\mathrm{IR}} \frac{\partial }{\partial \hat{g} } \left[ \hat{g} \mathcal{W}(\hat{g}) \right],
    \end{equation}
    while the critical exponent $\eta = d - 2 - 2 \eta_2'$ \cite{nogueiraDeconfinedQuantumCriticality2008} characterizing the decay of the two-spin correlation function in three dimensions is given by
    \begin{equation}\label{eq:eta}
        \eta(\hat{g}) = 1 + \frac{\hat{u}_{\mathrm{IR}}}{4} - 2\hat{e}^2_{\mathrm{IR}} \frac{\partial }{\partial \hat{g} } \left[ \hat{g} \mathcal{W}(\hat{g}) \right].
    \end{equation}
    To one-loop order we moreover find that $\phi = 1 + \hat{u}_{\mathrm{IR}} N_b/16$, implying that the anisotropic part of the Néel correlation function dominates in the scaling limit.
    These critical exponents are plotted in Fig.~\ref{fig:critical_exponents}.
    Although quantum fluctuations tend to reduce the large mean-field prediction $\eta_{\mathrm{MFT}} = 1$ that follows from the deconfinement of spinons \cite{nogueiraDeconfinedQuantumCriticality2008}, we see that with large values of the topological mass, $\eta$ can even exceed $1$.
    A large value for $\eta$ is one of the universal signatures of spin fractionalization \cite{isakovUniversalSignaturesFractionalized2012,shytaDeconfinedCriticalityBosonization2021}.
    Lastly, in the limit $\hat{g}\to\infty$ where the spinons are completely deconfined we recover the exponents associated with the $\mathrm{O}(2N_b)$ universality class 
    \begin{equation}\label{eq:O2N}
	     \nu_{\mathrm{O}(2N_b)}^{-1} = 2 - \frac{N_b + 1}{N_b + 4}  \quad \text{\&} \quad \eta_{\mathrm{O}(2N_b)}= 1 + \frac{2}{N_b + 4}.
    \end{equation}
    
    \subsubsection{Regime $D < 0$}
    
    Outside the critical regime, we expect runaway flow to negative $\hat{u}$ which begs for the inclusion of $\abs{z}^6$ terms in the Landau theory and therefore implies a first-order transition \cite{halperinFirstOrderPhaseTransitions1974}.
    However, approaching the critical line, criticality reappears with highly nontrivial scaling \cite{kaplanConformalityLost2009,nogueiraConformalityLossQuantum2019}. 
    Integrating the $\beta$ function in the regime $D<0$ in the asymptotic limit $\abs{\Delta}\to 0$ now yields 
    \begin{equation}
        \xi \sim \exp \left( \frac{8}{N_b+4} \frac{\pi}{\abs{\Delta}} \right),
    \end{equation}
    for all $u < u_{*}$.
    Recalling that $\abs{\Delta} \sim \abs{\sqrt{D(\hat{g},N_b)}}$ this implies Berezinskii--Kosterlitz--Thouless (BKT) scaling if $D$ is linear in $g-g_c$ close to the critical line \cite{kosterlitzOrderingMetastabilityPhase1973a,nelsonUniversalJumpSuperfluid1977}, also referred to as ``walking" behavior in this context \cite{gorbenkoWalkingWeakFirstorder2018}.
    Indeed, for $N_b = 2$ we find $\hat{g}_c \simeq 1.795$ and $D(\hat{g}) = D_0(\hat{g} - \hat{g}_c)$ with $D_0 \simeq 5.845$.
    For $\hat{u} > \hat{u}_{*}$ we find 
    \begin{equation}
        \xi \sim \exp \left( \frac{8}{N_b + 4} \frac{1}{\hat{u} - \hat{u}_{*}} \right),
    \end{equation}
    implying a first-order transition. 
    However, if $\hat{u}$ is close to $\hat{u}_{*}$ the correlation length is large but finite, which is a sign of deconfined pseudocriticality \cite{maTheoryDeconfinedPseudocriticality2020}.
    
	\section{Chiral Symmetry breaking}\label{sec:csb}
    
    The analysis of the previous section elucidates the effect of the spin liquid on the DQCP.
    An important issue left to consider is how the bosonic spinon theory affects the spin liquid.
    Indeed, while the omission of local matter couplings is justified in an RG sense, such an argument is based on the Dirac fermions being well-defined excitations.
    Hence it implicitly assumes the spin liquid is stable.
    To investigate whether this assumption still holds after we couple the layers together, we ask whether the system displays an increased predisposition to chiral symmetry breaking (CSB) with finite $\hat{g}$. 
    CSB in the theory describing a $\mathrm{U}(1)$ spin liquid is associated with the onset of antiferromagnetic order, which indicates spinon confinement even for a non-compact gauge theory \cite{kimTheorySpinExcitations1999b,nogueiraDeconfinedQuantumCriticality2008}

    Let the inverse of the dressed fermionic propagator be given by $G^{-1}(p) = \iu \slashed{p} Z(p) + \Sigma(p)$, where $Z(p)$ is the wavefunction renormalization and $\Sigma(p)$ is a dynamical mass \cite{appelquistSpontaneousChiralsymmetryBreaking1986}.
    Following the standard Schwinger-Dyson analysis of this problem \cite{appelquistSpontaneousChiralsymmetryBreaking1986,appelquistCriticalBehavior2+1Dimensional1988} we obtain the lowest-order self-consistent gap equation for $\Sigma(p)$ by assuming that $Z(p)=1$ and neglecting vertex corrections
    \begin{equation}
	   \Sigma(p) = f^2 \int \frac{\D^3 k}{(2\pi)^3} \frac{\Sigma(k)}{k^2 + \Sigma^2(k)} \sum_{\mu=1}^{3} \mathcal{D}^{11}_{\mu\mu}(p-k),
    \end{equation}
    where $\mathcal{D}^{11}_{\mu\nu}(k)$ is the dressed propagator of the $a_{\mu}$ gauge field.
    Realizing that its vacuum polarization to one-loop order is only due to the fermion loop diagram in Fig.~\ref{subfig:b}, we obtain the gap equation for $\Sigma$ given by
		\begin{equation}\label{eq:Sigma_general}
			\begin{split}
				\Sigma(p) = 2 f^2 \int \frac{\D^3 k }{(2\pi)^3} \frac{\Sigma(k)}{k^2 + \Sigma^2(k)} & \\
				\times \frac{1}{(k-p)^2 + g^2} &\frac{\abs{k-p}}{\abs{k-p} + N_f f^2 / 8}.
			\end{split}
		\end{equation}
		Following Ref.~\cite{appelquistCriticalBehavior2+1Dimensional1988} we can assume that the gauge field propagator is dominated by its one-loop vacuum polarization provided that we supplement the integral over $k$ with a cutoff at $\Lambda \simeq N_f f^2/8$.
		Assuming that $\Sigma(k) = \Sigma(\abs{k})$ and doing the angular integral of Eq.~\eqref{eq:Sigma_general} yields
	\begin{widetext}
		\begin{equation}\label{eq:Sigma_approx}
			\Sigma(p) = \frac{4}{N_f \pi^2 p } \int_{0}^{\Lambda} \D k \frac{k \Sigma(k)}{k^2 + \Sigma^2(k)} \left( k + p - \abs{k - p} - g \left\{ \arctan\left(\frac{k+p}{g}\right) - \arctan\left(\frac{\abs{k-p}}{g}\right) \right\} \right),
		\end{equation}
	\end{widetext}
    where we have assumed without loss of generality that $g\geq0$.
    The effect of finite $g$ can be qualitatively understood from the integral equation in Eq.~\eqref{eq:Sigma_approx}.
    In the limit $g\to 0$ we recover the classical results of $\mathrm{QED}_3$, where it is well-known that a non-trivial solution $\Sigma(p)$ of the above equation only exists if $N_f < 32/\pi^2$ \cite{appelquistSpontaneousChiralsymmetryBreaking1986,appelquistCriticalBehavior2+1Dimensional1988}. 
    From Eq.~\eqref{eq:Sigma_approx} it is clear that $\partial \Sigma/\partial g < 0 $, meaning that finite $g$ diminishes $\Sigma$.
    In the limit $g/\Lambda \to \infty$ the quantity in the parenthesis vanishes and $\Sigma$ must vanish identically as well. 
    Thus, we see that a finite $g$ reduces the system's tendency towards CSB.
    Qualitatively, this can also be understood by the fact that CSB in $\mathrm{QED}_3$ is caused by the long-ranged fluctuations of the gapless gauge field.
    These fluctuations are softened by the presence of the BF term.
    
	\section{Discussion and conclusion}\label{sec:disc}

    We have studied the low-energy effective theory of a $\mathrm{U}(1)$ Dirac spin liquid coupled to a quantum antiferromagnet close to its Néel-VBS transition. 
	Using a fixed-dimension RG calculation, we have examined the critical behavior of the quantum antiferromagnet as a function of the coefficient $g$ of the topological BF term.
	A finite coupling $g$ increases the quantum antiferromagnet's propensity for a DQCP in the sense that it reduces the critical number of complex components of the spinon field required for the existence of a charged fixed point.	
	In particular, increasing $g$ beyond a critical value $g_c$ furnishes an infrared-stable fixed point and marks the point at which the imaginary part of the fixed points vanishes, followed by BKT-type scaling of the correlation length in $2+1$ spacetime dimensions. 
    Approaching $g_c$ from above leads to a fixed point collision and generically a first-order transition.
    This theory therefore realizes the paradigmatic scenario introduced by Kaplan \textit{et al.} to describe conformality loss \cite{kaplanConformalityLost2009}.
	In the critical domain, the correlation length and the two-spin correlation function are characterized by critical exponents that vary continuously with $g$.

    Previous studies of the quantum criticality of theories with coupled, charged fermions and bosons have mainly considered the case where the matter fields couple to the same $\mathrm{U}(1)$ gauge field \cite{kleinertCriticalBehaviorGinzburgLandau2002,nogueiraDeconfinedQuantumCriticality2008,kaulQuantumCriticalityGauge2008}.
    While this has similar effects on the DQCP, the BF-mediated coupling we have considered is fundamentally different in the sense that it realizes a new universality class and is relevant for bilayers of fractionalized spin systems.
    An interesting avenue for future research would be to provide an explicit microscopic derivation of the BF term in the context of spin-liquid bilayer systems.  
	Moreover, since the AHM appears as an effective description of a variety of condensed matter systems, our study is likely to be relevant in contexts beyond the present one.
	
	\begin{acknowledgments}
		We acknowledge support from the Norwegian Research Council through Grant No. 262633, ``Center of Excellence on Quantum Spintronics” and Grant No. 323766, as well as COST Action CA21144  ``Superconducting Nanodevices and Quantum Materials for Coherent Manipulation".
	\end{acknowledgments}
	
	\appendix
	
	\section{Irrelevance of local spin-spin interaction}\label{app:biquadratic}

    Let us elaborate on the rationale for omitting local matter couplings.
    To write such a coupling requires a microscopic theory for the spin liquid, which usually entails constructing a parton gauge theory with an associated mean-field ansatz minimizing the free energy \cite{wenMeanfieldTheorySpinliquid1991b,wenQuantumFieldTheory2007}.
    To simplify this discussion we therefore consider the effect of the simplest kind of local matter interaction between the layers that is consistent with it originating from a spin-spin interaction, and for the moment neglect the $a_{\mu}$ gauge field. 
    This takes the form 
    \begin{equation}\label{eq:spinspin}
        \calL_{\mathrm{int}} = \lambda z^{\dagger} z \bar{\psi} \psi.
    \end{equation}
    The constraint of considering terms that may originate from a spin-spin interaction precludes all Yukawa-type couplings, which 
    relevantly modify the RG flow of the AHM \cite{royQuantumSuperconductingCriticality2013,jianFermioninducedQuantumCritical2017,torresFermioninducedQuantumCriticality2018,yinFermioninducedDynamicalCritical2021,schererGaugefieldassistedKekuleQuantum2016,classenFluctuationinducedContinuousTransition2017,boyackTransitionAlgebraic$mathbbZ_2$2018,zerfCriticalPropertiesValencebondsolid2020,seifertFractionalizedFermionicQuantum2020,rayFractionalizedQuantumCriticality2021}.
    
    That the matter interaction in Eq.~\eqref{eq:spinspin} does not have the same effect as a Yukawa term can be qualitatively understood by integrating out the fermions and the $A_{\mu}$ gauge field under the assumption of spatially uniform $z$.
    Under this assumption, both functional integrals are Gaussian and can be done exactly. 
    Integrating out $A$ produces a non-analytic term in $\abs{z}^2$ \cite{halperinFirstOrderPhaseTransitions1974} given by 
    \begin{align}
        \calL_{\mathrm{G}} &= \frac{1}{2}  \times 2 \int \frac{\D^3 k}{(2\pi)^3} \log\left( \v{k}^2 + 2 e^2 \abs{z}^2\right) \notag \\
        &= -\frac{\sqrt{2} e^3}{3\pi} \abs{z}^3,
    \end{align}
    where the factor of $2$ accounts for the two transverse degrees of freedom.
    This is the first sign of a fluctuation-induced first-order transition in the AHM.

    Now, integrating out the fermions yields the correction
    \begin{align}
        \calL_{\mathrm{F}} &= - \frac{1}{V} \tr \log( \slashed{\partial} + \lambda \abs{z}^2 ) \notag \\
        &= - 2 \int \frac{\D^3 k}{(2\pi)^3} \log \left( \v{k}^2 + \lambda^2 \abs{z}^4 \right) = + \frac{\lambda^3}{3\pi} \abs{z}^6.
    \end{align}
    The one--loop-corrected Landau theory for $z$ receives a $\abs{z}^6$ term from the fermion determinant, which is irrelevant as far as altering the fluctuation-induced first-order transition from the gauge field is concerned \cite{zinn-justinQuantumFieldTheory2021}. 
    In light of this, one might be tempted to consider a coupling between $z$ and $\psi$ of the Yukawa type, in which case the fermion determinant would yield a term $\sim \abs{z}^3$ with a positive sign, which can neutralize the effect of the gauge field fluctuations.
    While theories with this kind of interaction find applications in a variety of systems, the Yukawa coupling is not appropriate for describing systems in which the bosons and fermions emerge from independent degrees of freedom. 
    
    \onecolumngrid
	
	\section{Details of the RG calculation}\label{app:RG}

    In this appendix, we provide details on the RG calculation used to obtain the $\beta$ functions in Eq.~
	\eqref{eq:beta_functions_3}.
	The procedure parallels that of refs.~\cite{herbutCriticalFluctuationsSuperconductors1996,nogueiraFieldTheoreticApproaches2004b} for the RG of the AHM.
	The issue of obtaining a charged critical point of the AHM boils down to finding non-trivial, simultaneous solutions of the $\beta$ functions for $e^2$ and $u$.
	The calculations are therefore conveniently done in the massless limit $m^2 = 0$, while evaluating the Feynman diagrams at nonzero external momenta. 
	The magnitude of the momentum $q$ flowing through the graph sets the renormalization group scale $\mu$, and regulates all infrared divergences \cite{kleinertCriticalProperties$phi^4$2001}. 
	In Ref.~\cite{nogueiraFieldTheoreticApproaches2004b} it was shown that using this approach directly in $d=3$ dimensions (i.e., without resorting to dimensional regularization $d=4-\epsilon$) reduces the critical $N_b$ found by Halperin, Lubensky and Ma drastically \cite{halperinFirstOrderPhaseTransitions1974}, suggesting that a fixed-dimension approach to this problem is more controlled \cite{kleinertChargedFixedPoint2003}.
    
    \begin{fmffile}{small_diagrams}
		\fmfset{dot_size}{1.5pt}
		\fmfset{thin}{0.6pt}
		\fmfset{wiggly_len}{3pt}
		\fmfset{dash_len}{3pt}
		In the following, we compute the diagrams listed in Fig.~\ref{fig:feynman_diagrams_1}.
		Let us first consider those that do not involve internal gauge bosons.
        The first two provide the gauge-field renormalization constants.
		\begin{align}
			\eqgraph{1.5mm}{0ex}{\fmfreuse{ZA}}
			&=
			- \frac{e^2}{2} S_{\eqgraph{0mm}{0ex}{\begin{fmfgraph*}(15,7)
						\fmfkeep{diag_a}
						\fmfleft{i}
						\fmfright{r}
						\fmf{boson,tension=1}{i,v1}
						\fmf{plain,tension=0.4,right=0.8}{v1,v2}
						\fmf{plain,tension=0.4,right=0.8}{v2,v1}
						\fmf{boson,tension=1}{v2,r}
						\fmfdot{v1,v2}
			\end{fmfgraph*}}} \int \frac{\D^d k}{(2\pi)^d} \frac{(q-2k)_{\mu} (q-2k)_{\nu}}{k^2 (k-q)^2} \notag
			\\
			&= - \frac{e^2}{2} S_{\eqgraph{0mm}{0ex}{\fmfreuse{diag_a}}} \int_{0}^{1} \D s \int \frac{\D^d k}{(2\pi)^d} \frac{(q-2k)_{\mu} (q-2k)_{\nu}}{((k-q s)^2 + q^2 s(1-s))^2} \notag \\
			&=  - \frac{e^2}{2} S_{\eqgraph{0mm}{0ex}{\fmfreuse{diag_a}}} \int_{0}^{1} \D s \int \frac{\D^d \ell }{(2\pi)^d} \left[ (1-2s)^2 \frac{q_{\mu} q_{\nu}}{(\ell^2 + \Delta(s))^2} + 4 \frac{\ell_{\mu} \ell_{\nu}}{(\ell^2 + \Delta(s))^2} \right],
		\end{align}
		where we have introduced $\Delta(s) \coloneqq q^2 s (1-s)$, and changed variables to $\ell \coloneqq k - s q$ and neglected all terms linear in $\ell$ in the integrand. Doing the $\ell$ integral and subsequently the $s$ integral yields
		\begin{align}
			\eqgraph{1.5mm}{0ex}{\fmfreuse{ZA}} &= \hat{e}^2 S_{\eqgraph{0mm}{0ex}{\fmfreuse{diag_a}}}  \frac{1}{(4\pi)^{d/2}} \Gamma\left(1 - \frac{d}{2}\right) \frac{\Gamma^2(d/2)}{\Gamma(d)} \left( q_{\mu} q_{\nu} - \delta_{\mu\nu} q^2 \right).
		\end{align} 
		Hence, we define the renormalized $A$ field by choosing
		\begin{equation}
			(Z_{A} -1 ) = 2 S_{\eqgraph{0mm}{0ex}{\fmfreuse{diag_a}}} \hat{e}^2 F(d),
		\end{equation}
		where we have defined 
		\begin{equation}\label{eq:F}
			F(d) \equiv \frac{1}{(4\pi)^{d/2}} \Gamma\left(1 - \frac{d}{2}\right) \frac{\Gamma^2(d/2)}{\Gamma(d)}.
		\end{equation}
		The symmetry factor of the diagram is $S_{\eqgraph{0mm}{0ex}{\fmfreuse{diag_a}}} = N_b$.

        The computation of the fermionic vacuum polarization similarly reads 
        \begin{align}
        	\eqgraph{1.5mm}{0ex}{\fmfreuse{Za}}&=
        	+ \frac{f^2}{2} S_{\eqgraph{0mm}{0ex}{\begin{fmfgraph*}(15,7)
						\fmfkeep{diag_b}
						\fmfleft{i}
						\fmfright{r}
						\fmf{boson,tension=1}{i,v1}
						\fmf{dashes,tension=0.4,right=0.8}{v1,v2}
						\fmf{dashes,tension=0.4,right=0.8}{v2,v1}
						\fmf{boson,tension=1}{v2,r}
						\fmfdot{v1,v2}
			\end{fmfgraph*}}} \int \frac{\D^d k}{(2\pi)^d} \frac{\iu^2}{k^2 (k-q)^2} \tr\left(\slashed{k} \iu\gamma_{\mu} (\slashed{k}-\slashed{q})\iu\gamma_{\nu}\right) \notag
        	\\
        	&= - \frac{f^2}{2} S_{\eqgraph{0mm}{0ex}{\fmfreuse{diag_b}}} \int \frac{\D^d k}{(2\pi)^d} \frac{-4}{k^2 (k-q)^2} \left[ k_{\mu}(k-q)_{\nu} + k_{\nu}(k - q)_{\mu} - \delta_{\mu\nu} k \cdot (k-q) \right] \notag \\
        	&=+ 2f^2 S_{\eqgraph{0mm}{0ex}{\fmfreuse{diag_b}}} \int_{0}^{1} \D s \int \frac{\D^d \ell }{(2\pi)^d} \frac{2 \ell_{\mu} \ell_{\nu} + 2 q_{\mu} q_{\nu} s(s-1) - \delta_{\mu\nu} \left( \ell^2  + q^2 s(s-1)\right) }{(\ell^2 + \Delta(s))^2} \notag \\
        	&= +4\hat{f}^2 S_{\eqgraph{0mm}{0ex}{\fmfreuse{diag_b}}} q^2\left[ \delta_{\mu\nu} -\frac{q_{\mu} q_{\nu}}{q^2} \right]\left(1 - \frac{d}{2}\right) F(d).
        \end{align}
        The symmetry factor of this diagram is $S_{\eqgraph{0mm}{0ex}{\fmfreuse{diag_b}}} = N_f$, so this yields 
        \begin{equation}
            (Z_{a} - 1) = 4 N_f \hat{f}^2 (d-2) F(d).
        \end{equation}
        
        Let us now compute the renormalization of $u$. 
		The contributing diagrams are the three last ones listed in Fig.~\ref{fig:feynman_diagrams_1}.
        The simplest one reads
		\begin{align}
			\eqgraph{1.5mm}{0ex}{\fmfreuse{Zu1}}
			&= -\frac{1}{2} S_{\eqgraph{0.0mm}{0ex}{\begin{fmfgraph*}(15,7)
						\fmfkeep{diag_c}
						\fmfleft{l1,l2}
						\fmfright{r1,r2}
						\fmf{plain}{l1,v1,l2}
						\fmf{plain,right=0.7,tension=0.5}{v1,v2,v1}
						\fmf{plain}{r1,v2,r2}
						\fmfdot{v1,v2}
			\end{fmfgraph*}}} \left(\frac{u}{2}\right)^2 \int \frac{\D^d k}{(2\pi)^d} \frac{1}{k^2(k-q)^2} \notag \\
			&=  -\frac{1}{2} S_{\eqgraph{0.0mm}{0ex}{\fmfreuse{diag_c}}} \frac{\hat{u} u}{4} \mu^{4-d} \int_{0}^{1}\D s \int \frac{\D^d \ell}{(2\pi)^d} \frac{1}{(\ell^2 + \Delta(s))^2} \notag \\
			&\equiv - \frac{1}{8} S_{\eqgraph{0.0mm}{0ex}{\fmfreuse{diag_c}}} \hat{u} u G(d),
		\end{align}
		where
		\begin{equation}\label{eq:G}
			G(d) \equiv \frac{1}{(4\pi)^{d/2}} \Gamma\left(2 - \frac{d}{2}\right) \frac{\Gamma^2\left(d/2 - 1\right)}{\Gamma(d-2)} \equiv 4 \pi F(d-2).
		\end{equation}
		The symmetry factor is given by $S_{\eqgraph{0.0mm}{0ex}{\fmfreuse{diag_c}}} = 4(N_b+4)$.

        What remains are diagrams involving the gauge field propagator. 
        These computations deviate from those in refs.~\cite{herbutCriticalFluctuationsSuperconductors1996,nogueiraFieldTheoreticApproaches2004b} due to the BF term.
        By temporarily introducing gauge-fixing terms in the Lagrangian, we can invert the inverse propagator $\mathcal{D}^{-1}_{\mu\nu}$ of the vector of gauge fields $\mathcal{A}_{\mu} \coloneqq \left( A_{\mu} \, a_{\mu} \right)^{\mathsf{T}}$.
        After doing so and fixing the Landau gauge, we can express the Fourier transform of the propagator by
        \begin{equation}
            \mathcal{D}_{\mu\nu}(k) = \frac{1}{k^2 + g^2} \begin{pmatrix}
                P_{\mu\nu}^{\perp}(k) & - g \epsilon_{\mu\nu\rho} k_{\rho} /k^2 \\
                - g \epsilon_{\mu\nu\rho} k_{\rho} /k^2 & P^{\perp}_{\mu\nu}(k)
            \end{pmatrix},
        \end{equation}
        where $P^{\perp}_{\mu\nu}(k) \equiv \delta_{\mu\nu} - k_{\mu} k_{\nu}/k^2 $ is the transverse projector.

        Let us compute the gauge field bubble integral
        \begin{align}
    	\eqgraph{1.5mm}{0ex}{\fmfreuse{Zu2}}
    	&= - \frac{1}{2} S_{\eqgraph{0.0mm}{0ex}{\begin{fmfgraph*}(15,7)
						\fmfkeep{diag_d}
						\fmfleft{l1,l2}
						\fmfright{r1,r2}
						\fmf{plain}{l1,v1,l2}
						\fmf{boson,right=0.7,tension=0.5}{v1,v2,v1}
						\fmf{plain}{r1,v2,r2}
						\fmfdot{v1,v2}
			\end{fmfgraph*}}} (e^2)^2 \int \frac{\D^d k }{(2\pi)^d} \frac{1}{(k^2 + g^2 ) \left(  (k-q)^2 + g^2 \right)} \left[ \left( \delta_{\mu\nu} - \frac{k_{\mu}k_{\nu}}{k^2} \right) \left( \delta_{\mu\nu} - \frac{(k-q)_{\mu}(k-q)_{\nu}}{(k-q)^2} \right) \right] \notag  \\
    	&= - \frac{1}{2} S_{\eqgraph{0.0mm}{0ex}{\fmfreuse{diag_d}}} (e^2)^2 \bigg[  (d-2) \int \frac{\D^d k }{(2\pi)^d} \frac{1}{(k^2 + g^2 ) \left(  (k-q)^2 + g^2 \right)} \notag \\ 
        &\hspace{7cm} + \int \frac{\D^d k }{(2\pi)^d} \frac{(k\cdot(k-q))^2}{k^2 (k-q)^2(k^2 + g^2 ) \left(  (k-q)^2 + g^2 \right)} \bigg]. \label{eq:bubble_integrals}
        \end{align}
        The strategy for computing this is to employ tricks of the same type as in the appendix of Ref.~\cite{nogueiraConformalityLossQuantum2019} to simplify the integrals in terms of the three ``one-loop" integrals of the form
        \begin{subequations}
        	\begin{align}
        		I_1(q) &\coloneqq \int \frac{\D^d k }{(2\pi)^d} \frac{1}{k^2 (k-q)^2} \\
        		I_2(q,g) &\coloneqq \int \frac{\D^d k }{(2\pi)^d} \frac{1}{(k^2 + g^2)(k-q)^2} \\
        		I_3(q,g) &\coloneqq \int \frac{\D^d k }{(2\pi)^d} \frac{1}{(k^2 + g^2)\left( (k-q)^2 + g^2 \right)}, 
        	\end{align}
        \end{subequations}
        which are simple to express in $d=3$.
        Note that the integrals $I_2$ and $I_3$ are well-defined even in the limit $q\to0$.

        To simplify the second integral of Eq.~\eqref{eq:bubble_integrals}, we use the rewriting
        \begin{align}
        	k \cdot(k-q) = \frac{1}{2} \left[ (k-q)^2 +k^2 - q^2 \right] = \frac{1}{2} \left[ (k-q)^2 + g^2  +k^2 + g^2 - q^2 \right] - g^2.
        \end{align}
        This yields six new integrals, which in turn can be reduced to a combination of the one-loop integrals by the repeated use of partial-fraction decompositions.
        By using the symmetry factor $S_{\eqgraph{0.0mm}{0ex}{\fmfreuse{diag_d}}} = 2$ and  defining $\mathcal{B}$ precisely as 
        \begin{equation}
	       -e^2 \hat{e}^2 \mathcal{B}(\hat{g};d) \coloneqq \eqgraph{1.5mm}{0ex}{\fmfreuse{Zu2}},
        \end{equation}
        and noting that all the one-loop integrals have $q$ and $\hat{g}$ dependence of the form $
        	I_i(q,g) \coloneqq q^{d-4} \mathcal{I}_{i}(\hat{g}),
        $
        some straightforward but tedious algebra yields
        \begin{align}
        	\mathcal{B}(\hat{g};d) = \frac{1}{4\hat{g}^4} \mathcal{I}_1(d) + \frac{1}{2} \mathcal{I}_{2,0}(\hat{g};d) 
        	- \left( \frac{1}{2} + \frac{1}{\hat{g}^2} + \frac{1}{2\hat{g}^4} \right)\mathcal{I}_2(\hat{g};d) + \left( (d-1) + \frac{1}{\hat{g}^2} + \frac{1}{4\hat{g}^2} \right) \mathcal{I}_3(\hat{g};d),
        \end{align}
        where $I_2(0,g) = q^{d-4} \mathcal{I}_{2,0}(\hat{g})$.
        In $d=3$ we find that 
        \begin{subequations}
        	\begin{align}
        		&\mathcal{I}_1 = \frac{1}{8} 
        		  &&\mathcal{I}_2(\hat{g}) = \frac{1}{4\pi} \arctan\left( \frac{1}{\abs{\hat{g}}}\right) \\
        		&\mathcal{I}_{2,0}(\hat{g}) = \frac{1}{4\pi \abs{\hat{g}}} 
        		&&\mathcal{I}_3(\hat{g}) =\frac{1}{4\pi} \arctan\left( \frac{1}{2 \abs{\hat{g}}}\right),
        	\end{align}
        \end{subequations}
        which yields exactly Eq.~\eqref{eq:B} of the main text.
        Together with the graph in Fig.~\ref{subfig:d} we can fix the renormalization constant of $u$ as
        \begin{equation}
            (Z_u - 1) = (N_b + 4) \hat{u} G(d) + 2 \hat{e}^2 \frac{e^2}{u} \mathcal{B}(\hat{g};d).
        \end{equation}
        It can be shown that taking the limit $\hat{g}\to 0$ with $d$ unspecified yields 
        \begin{equation}
            \mathcal{B}(0;d) = \frac{d}{4} (d-1) G(d).
        \end{equation}
        The factor $d/4$ was missing in the calculations of Ref.~\cite{nogueiraFieldTheoreticApproaches2004b}, leading to a slight overestimate of $N_b$ in $d=3$.

        Let us compute the wavefunction renormalization. 
        We define $\mathcal{W}$ precisely as 
        \begin{equation}
            -\hat{e}^2 q^2  \mathcal{W}(\hat{g};d) \coloneqq \eqgraph{1.5mm}{0ex}{\fmfreuse{Zz}}.
        \end{equation}
        The diagram reads
        \begin{align}
        	\eqgraph{1.5mm}{0ex}{\fmfreuse{Zz}}
        	&= - \frac{1}{2} S_{\eqgraph{0.0mm}{0ex}{\begin{fmfgraph*}(15,7)
						\fmfkeep{diag_c}
						\fmfleft{i}
						\fmfright{r}
						\fmf{plain}{i,v1}
						\fmf{plain,tension=0.3}{v1,v2}
						\fmf{plain}{v2,r}
						\fmf{boson,left=1,tension=0.3}{v1,v2}
						\fmfdot{v1,v2}
			\end{fmfgraph*}}} \, e^2 \int \frac{\D^d k}{(2\pi)^d} \left( k - 2 q \right)_{\mu} \frac{1}{k^2 + g^2}\left( \delta_{\mu\nu} - \frac{k_{\mu} k_{\nu}}{k^2}\right) (k- 2 q)_{\nu}\frac{1}{(k-q)^2} \notag \\
        	&= - \frac{1}{2} S_{\eqgraph{0.0mm}{0ex}{\fmfreuse{diag_c}}} \,  e^2 \int \frac{\D^d k}{(2\pi)^d} \frac{1}{(k^2+g^2) (k-q)^2} \left[ (k-2q)^2  -  (k-2q)_{\mu} \frac{k_{\mu} k_{\nu}}{k^2} (k-2q)_{\nu}  \right] \notag \\
        	&= - 2 S_{\eqgraph{0.0mm}{0ex}{\fmfreuse{diag_c}}} \, e^2 \int \frac{\D^d k}{(2\pi)^d} \left[ \frac{q^2}{(k^2+g^2) (k-q)^2} - q_{\mu} q_{\nu} \frac{k_{\mu} k_{\nu}}{k^2 (k^2+ g^2)(k-q)^2} \right].
        \end{align}
        The first diagram can be expressed in terms of $I_2(q,g)$.
        The second one can be simplified using the same methods as above. 
        Using the symmetry factor $S_{\eqgraph{0.0mm}{0ex}{\fmfreuse{diag_c}}}= 2$ we find in $d=3$ that
        \begin{equation}\label{eq:W_app}
        	\mathcal{W}(\hat{g}) = \frac{1}{\pi} \left[ 1 + \frac{1}{4\hat{g}^2} \left(\hat{g}^2 - 1\right)^2 \right] \arctan\left( \frac{1}{\abs{\hat{g}}}\right) - \frac{1}{8\hat{g}^2} - \frac{1}{4\pi} \left(\hat{g}^2 - 1\right)\frac{1}{\abs{\hat{g}}}.
        \end{equation}
        This yields the wavefunction renormalization 
        \begin{equation}
            (Z_z - 1) = \hat{e}^2 \mathcal{W}(\hat{g};d)
        \end{equation}
        Taking the limit $\hat{g}\to 0$ with $d$ unspecified yields $\mathcal{W}(0;d) = (d-1) G(d)$.
    \end{fmffile}

    \section{Critical exponents}\label{app:critical_exponents}

    Lastly, let us sketch the computation of $Z_2$ and $Z_2'$ leading to the critical exponents $\nu$ and $\eta$.
    In the massless theory, we compute these renormalization factors by insertions of the sum and difference of the operators $\abs{z_1}^2$ and $\abs{z_2}^2$ in the correlation function $\langle \bar{z}_{1}(x)z_1(0)\rangle$ \cite{nogueiraDeconfinedQuantumCriticality2008,zinn-justinQuantumFieldTheory2021}.
    To facilitate this, we add to the Lagrangian the source terms $m_1^2 \abs{z_1}^2 + m_2^2 \abs{z_2}^2$ and perform derivatives with respect to $m_1^2$ and $m_2^2$ of the correlation function $\langle \bar{z}_{1}(x)z_1(0)\rangle$ to one-loop order. 
    The resulting diagrams form a subset of those renormalizing the four-point function and are considered as such. 
    The symmetry and weight factors associated with these diagrams are however inherited from the original two-point function \cite{kleinertCriticalProperties$phi^4$2001}.

    \begin{fmffile}{crossover}
	\fmfset{dash_len}{1.5mm}
	\fmfset{wiggly_len}{2mm}
	\fmfset{dot_size}{3pt}
    By illustrating $z_1$ as a black plain line and $z_2$ as a gray plain line, we can diagrammatically represent the renormalization factors as 
    \begin{align}
        Z_2 &= -\frac{\partial}{\partial m_1^2} \left( 
			\eqgraph{1.5mm}{0ex}
			{\begin{fmfgraph*}(30,10)
					\fmfleft{i}
					\fmfright{r}
					\fmf{plain}{i,v,r}
			\end{fmfgraph*}}
			+
			\eqgraph{1.5mm}{0ex}{
			\begin{fmfgraph*}(30,20)
				\fmfleft{i}
				\fmfright{r}
				\fmf{plain}{i,v,r}
				\fmf{plain,right=0.7,tension=0.7}{v,v}
				\fmfdot{v}
			\end{fmfgraph*}}
			+
			\eqgraph{1.5mm}{0ex}{
			\begin{fmfgraph*}(30,20)
				\fmfleft{i}
				\fmfright{r}
				\fmf{plain}{i,v,r}
				\fmf{plain,right=0.7,tension=0.7,foreground=(0.7,,0.7,,0.7)}{v,v}
				\fmfdot{v}
			\end{fmfgraph*}}
			+
			\eqgraph{1.5mm}{0ex}{
			\begin{fmfgraph*}(30,20)
				\fmfleft{i}
				\fmfright{r}
				\fmf{plain}{i,v1}
				\fmf{plain,tension=0.3}{v1,v2}
				\fmf{plain}{v2,r}
				\fmf{boson,left=1,tension=0.3}{v1,v2}
				\fmfdot{v1,v2}
			\end{fmfgraph*}	
			}
		\right) \biggr\lvert_{m_1 = 0}
        -
        \frac{\partial}{\partial m_2^2} \left( 
		\eqgraph{1.5mm}{0ex}
		{\begin{fmfgraph*}(30,10)
				\fmfleft{i}
				\fmfright{r}
				\fmf{plain}{i,v,r}
		\end{fmfgraph*}}
		+
		\eqgraph{1.5mm}{0ex}{
			\begin{fmfgraph*}(30,20)
				\fmfleft{i}
				\fmfright{r}
				\fmf{plain}{i,v,r}
				\fmf{plain,right=0.7,tension=0.7}{v,v}
				\fmfdot{v}
		\end{fmfgraph*}}
		+
		\eqgraph{1.5mm}{0ex}{
			\begin{fmfgraph*}(30,20)
				\fmfleft{i}
				\fmfright{r}
				\fmf{plain}{i,v,r}
				\fmf{plain,right=0.7,tension=0.7,foreground=(0.7,,0.7,,0.7)}{v,v}
				\fmfdot{v}
		\end{fmfgraph*}}
		+
		\eqgraph{1.5mm}{0ex}{
			\begin{fmfgraph*}(30,20)
				\fmfleft{i}
				\fmfright{r}
				\fmf{plain}{i,v1}
				\fmf{plain,tension=0.3}{v1,v2}
				\fmf{plain}{v2,r}
				\fmf{boson,left=1,tension=0.3}{v1,v2}
				\fmfdot{v1,v2}
			\end{fmfgraph*}	
		}
		\right) \biggr\lvert_{m_2 = 0} \notag \\
        &= 
        \eqgraph{1.5mm}{0ex}
		{\begin{fmfgraph*}(30,10)
				\fmfleft{i}
				\fmfright{r}
				\fmftop{tl1,tl2,t1,t2,tr1,tr2}
				\fmf{plain}{i,v,r}
				\fmffreeze
				\fmf{plain,tension=5}{v,t1}
				\fmf{plain,tension=5}{v,t2}
				\fmfdot{v}
		\end{fmfgraph*}}
		+
		\eqgraph{1.5mm}{0ex}
		{\begin{fmfgraph*}(30,30)
				\fmfkeep{Gamma1}
				\fmfleft{i}
				\fmfright{r}
				\fmftop{tl1,tl2,t1,t2,tr1,tr2}
				\fmf{plain}{i,v,r}
				\fmffreeze
				\fmf{plain,right=1,tension=0.7}{v,vt,v}
				\fmf{plain,tension=5}{vt,t1}
				\fmf{plain,tension=5}{vt,t2}
				\fmfdot{v,vt}
		\end{fmfgraph*}}
		+
		\eqgraph{1.5mm}{0ex}
		{\begin{fmfgraph*}(30,20)
				\fmfleft{i}
				\fmfright{r}
				\fmftop{tl1,tl2,t1,t2,tr1,tr2}
				\fmf{plain}{i,v1}
				\fmf{wiggly,tension=0.7}{v1,v2}
				\fmf{plain}{v2,r}
				\fmffreeze
				\fmf{plain,left=0.5,tension=0.7}{v1,vt,v2}
				\fmf{plain,tension=5}{vt,t1}
				\fmf{plain,tension=5}{vt,t2}
				\fmfdot{v1,v2,vt}
		\end{fmfgraph*}}
        +
        \eqgraph{1.5mm}{0ex}
		{\begin{fmfgraph*}(30,30)
				\fmfkeep{Gamma2}
				\fmfleft{i}
				\fmfright{r}
				\fmftop{tl1,tl2,t1,t2,tr1,tr2}
				\fmf{plain}{i,v,r}
				\fmffreeze
				\fmf{plain,right=1,tension=0.7,foreground=(0.7,,0.7,,0.7)}{v,vt,v}
				\fmf{plain,tension=5,foreground=(0.7,,0.7,,0.7)}{vt,t1}
				\fmf{plain,tension=5,foreground=(0.7,,0.7,,0.7)}{vt,t2}
				\fmfdot{v,vt}
		\end{fmfgraph*}},
    \end{align}
    and similarly 
    \begin{equation}
        Z_2' =  \eqgraph{1.5mm}{0ex}
		{\begin{fmfgraph*}(30,10)
				\fmfleft{i}
				\fmfright{r}
				\fmftop{tl1,tl2,t1,t2,tr1,tr2}
				\fmf{plain}{i,v,r}
				\fmffreeze
				\fmf{plain,tension=5}{v,t1}
				\fmf{plain,tension=5}{v,t2}
				\fmfdot{v}
		\end{fmfgraph*}}
		+
		\eqgraph{1.5mm}{0ex}
		{\begin{fmfgraph*}(30,30)
				\fmfkeep{Gamma1}
				\fmfleft{i}
				\fmfright{r}
				\fmftop{tl1,tl2,t1,t2,tr1,tr2}
				\fmf{plain}{i,v,r}
				\fmffreeze
				\fmf{plain,right=1,tension=0.7}{v,vt,v}
				\fmf{plain,tension=5}{vt,t1}
				\fmf{plain,tension=5}{vt,t2}
				\fmfdot{v,vt}
		\end{fmfgraph*}}
		+
		\eqgraph{1.5mm}{0ex}
		{\begin{fmfgraph*}(30,20)
				\fmfleft{i}
				\fmfright{r}
				\fmftop{tl1,tl2,t1,t2,tr1,tr2}
				\fmf{plain}{i,v1}
				\fmf{wiggly,tension=0.7}{v1,v2}
				\fmf{plain}{v2,r}
				\fmffreeze
				\fmf{plain,left=0.5,tension=0.7}{v1,vt,v2}
				\fmf{plain,tension=5}{vt,t1}
				\fmf{plain,tension=5}{vt,t2}
				\fmfdot{v1,v2,vt}
		\end{fmfgraph*}}
        -
        \eqgraph{1.5mm}{0ex}
		{\begin{fmfgraph*}(30,30)
				\fmfkeep{Gamma2}
				\fmfleft{i}
				\fmfright{r}
				\fmftop{tl1,tl2,t1,t2,tr1,tr2}
				\fmf{plain}{i,v,r}
				\fmffreeze
				\fmf{plain,right=1,tension=0.7,foreground=(0.7,,0.7,,0.7)}{v,vt,v}
				\fmf{plain,tension=5,foreground=(0.7,,0.7,,0.7)}{vt,t1}
				\fmf{plain,tension=5,foreground=(0.7,,0.7,,0.7)}{vt,t2}
				\fmfdot{v,vt}
		\end{fmfgraph*}},
    \end{equation}
    where the short amputated lines represent an additional four-point vertex.
    The third diagram appearing in these equations vanishes in the Landau gauge. 
    The result is 
    \begin{equation}
        (Z_2 - 1) = \Bigg( \eqgraph{1.5mm}{0ex}{\fmfreuse{Gamma1}} + \eqgraph{1.5mm}{0ex}{\fmfreuse{Gamma2}} \Bigg) = (N_b + 1) \hat{u} G(d) \quad \text{\&} \quad (Z_2' - 1) = \Bigg( \eqgraph{1.5mm}{0ex}{\fmfreuse{Gamma1}} - \eqgraph{1.5mm}{0ex}{\fmfreuse{Gamma2}} \Bigg) =  \hat{u} G(d).
    \end{equation}
    
    \end{fmffile}
	\twocolumngrid
	
	\bibliography{references}
	
\end{document}